\documentclass[prl,aps,twocolumn,showpacs]{revtex4}
\usepackage{graphicx,amsmath,amsfonts,latexsym,bm,eso-pic}

\def \eps{\epsilon}
\def \beq{\begin{equation}}
\def \eeq{\end{equation}}
\def \beqa{\begin{eqnarray}}
\def \eeqa{\end{eqnarray}}
\def \sx{\sigma_x}

\def \sz{\sigma_z}
\def \tx{\tau_x}

\def \tz{\tau_z}

\newcommand{\rem}[1]{}

\def\sign{\mathrm{sign}}

 \def\nn{\nonumber}

\begin{document}

\title{Non-equilibrium Josephson effect through helical edge states}

\author{Driss M. Badiane} 
\author{Manuel Houzet}
\author{Julia S. Meyer}
\affiliation{SPSMS, UMR-E CEA / UJF-Grenoble 1, INAC, Grenoble, F-38054, France}
\date{\today}


%71.10.Pm	Fermions in reduced dimensions (anyons, composite fermions, Luttinger liquid, etc.) (for anyon mechanism in superconductors, see 74.20.Mn)
%74.45.+c	Proximity effects; Andreev reflection; SN and SNS junctions
%05.40.Ca	Noise
%03.67.Lx	Quantum computation architectures and implementations

\begin{abstract}
We study Josephson junctions between superconductors connected through the helical edge states of a two-dimensional topological insulator in the presence of a magnetic barrier. As the equilibrium Andreev bound states of the junction are $4\pi$-periodic in the superconducting phase difference, it was speculated that, at finite dc bias voltage, the junction exhibits a fractional Josephson effect with half the Josephson frequency. Using the scattering matrix formalism, we show that signatures of this effect can be seen in the finite-frequency current noise. Furthermore, we discuss other manifestations of the Majorana bound states forming at the edges of the superconductors.
\end{abstract}

\maketitle

Unlike ordinary insulators, topological insulators (TI) admit robust conducting states at their boundaries. These states display unique properties. For instance, a two-dimensional quantum spin-Hall insulator 
has helical edge states with up spins propagating in one direction and down spins propagating in the other direction \cite{hasan2010,qi2008}. Signatures of these helical edge states have been revealed in transport measurements on HgTe/CdTe \cite{konig2007} and InAs/GaSb \cite{knez2011} quantum well structures. 

A conventional superconductor (S) attached to such edge states induces topological superconductivity by the proximity effect. The resulting topological superconductor has been predicted to support zero-energy Majorana bound state (MBS) at an interface with a topologically trivial region \cite{kitaev2001}. Majorana fermions have attracted a lot of attention 
because they are promising for topologically protected quantum computation \cite{kitaev2003}. Indeed, a pair of spatially separated Majorana fermions form a Dirac fermion that could be used as a quantum bit. As the information is encoded non-locally, it would be quite immune to decoherence when the MBS are far away. 

When two topological superconductors are connected, 
a spectacular fractional Josephson effect has been predicted \cite{kitaev2001,kwon2003,fu2009}.
The Majorana bound states localized on either side of the junction hybridize and form an Andreev bound state with energy $\epsilon(\varphi)=\pm\sqrt{D}\Delta\cos(\varphi/2)$, where $\varphi$ is the phase difference between the two superconductors, $D$ is the transmission of the junction, and $\Delta$ is the superconducting gap. The energy $\epsilon(\varphi)$ is $4\pi$-periodic with respect to the superconducting phase difference, and, in the absence of inelastic processes, the crossing between the two states at $\varphi=\pi$ is protected by fermion parity \cite{fu2009}. Thus, if the phase is varied adiabatically, the system should remain in the same state. The resulting Josephson current would then be given by $I\sim\partial_\varphi\epsilon(\varphi)\propto\sin(\varphi/2)$. As a result, under dc bias voltage $V$, such a system has been predicted to manifest an ac Josephson effect at frequency $\omega_J/2=eV/\hbar$, that is half the Josephson frequency. 

The above prediction is an out-of-equilibrium property. Indeed, in equilibrium, the $2\pi$-periodicity of the Josephson relation is restored as the ground state corresponds to the energy level $\epsilon(\varphi)=-\sqrt{D}\Delta|\cos(\varphi/2)|$ yielding the Josephson current. To establish equilibrium an infinitesimal rate of inelastic scattering, e.g., due to the coupling with a bath is necessary. 
It is well known that, in conventional Josephson junctions under dc bias, inelastic scattering is induced by non-adiabatic transitions between Andreev bound states and the continuum of states above the gap.  Through a mechanism known as multiple Andreev reflections (MAR) \cite{klapwijk1982}, particles in the junction gain an energy $eV$ at each traversal until 
they have acquired sufficient energy to escape into the continuum. In particular, this manifests itself by a dissipative current flowing through the junction at subgap voltages. Thus, one may wonder how robust is the prediction of a fractional Josephson effect. The aim of the present work is to address this question by computing the current and noise of an S/TI/S junction. 

\begin{figure}
\includegraphics[width=1\linewidth]{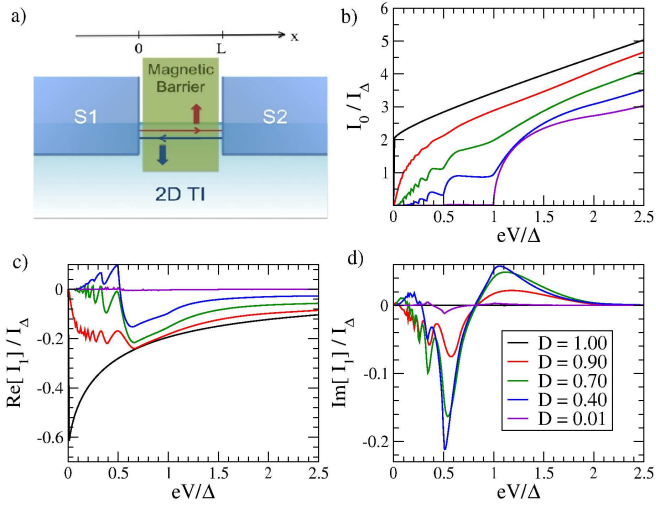}
\caption{
a) Schematic view of the S/TI/S junction. b) Dc current, c) real part, and d) imaginary part of the first harmonic of the ac current as a function of applied bias for various transparencies. Here $I_\Delta=G_N\Delta/e$ with $G_N=De^2/h$.}
\label{fig1} 
\vspace*{-0.3cm}
\end{figure}

The setup is shown in Fig.~\ref{fig1} \cite{otheredge}. To proceed, we adopt the Landauer-B\"uttiker formalism \cite{beenakker1991,bratus1995,averin1995}. The scattering states of the system are the eigenstates $\Phi^T=(u_+,v_+,u_-,v_-)$ -- where $u_\pm$ and $v_\pm$ are the electron and hole
components associated with right ($+$) and left ($-$) movers, respectively -- of the Bogoliubov-de Gennes (BdG) Hamiltonian
\beq
{\cal H}=vp \sz \tz+M(x) \sx+\Delta(x)e^{i\phi(x)\tz}\tx.
\eeq
Here, $v$ is the Fermi velocity, $p$ is the momentum operator, $M(x)=M\theta(x)\theta(L-x)$ is a transverse exchange field within the junction, $\Delta(x)=\Delta[\theta(-x)+\theta(x-L)]$ and $\phi(x)=\phi_l\theta(-x)+\phi_r\theta(x-L)$ are the amplitude and phase of the order parameter in the left and right leads ($\varphi=\phi_l-\phi_r$), and $\sigma_i$, $\tau_j$ ($i,j=x,y,z$) are Pauli matrices acting on the spin (equivalent to $+/-$) and Nambu ($u/v$) spaces, respectively. All energies are measured from the chemical potential. 

The current operator at the junction in terms of electron operators of right and left movers reads
$\hat I=ev[\hat\psi_+^\dagger(0)\hat\psi_+(0)-\hat\psi_-^\dagger(0)\hat\psi_-(0)]$. It can be expressed through the scattering states by using a Bogoliubov transformation, $\hat\psi_s(x)
=
\sum_\nu
\left[
u_{s\nu}(x)
\hat\gamma_\nu
-s v^*_{-s\nu}(x)
\hat\gamma^\dagger_\nu
\right]$,
where $s=\pm$ and $\nu=\{\eps,i,\alpha\}$ labels an incoming state with positive energy $\eps$, from the lead $i=l,r$, and of the type $\alpha=e,h$.

Due to multiple Andreev reflections \cite{klapwijk1982}, scattering states correspond to a superposition of states with energy $\epsilon+2neV$ ($n$ integer). For instance, the wave function of an incoming electron ($e$) with energy $\eps$ from the left ($l$) lead can be written in the form
\begin{subequations}
\label{eq:wavefunction}
\beq
\Phi_{\eps}^{el}(0,t)=J\sum_n\left(\begin{array}{c}
\delta_{n0}+a_{2n}A_n \\
A_n \\
B_n \\
a_{2n} B_n
\end{array}\right)
e^{-i(\eps+2neV)t}
\eeq
and
\beqa
\Phi_{\eps}^{el}(L,t)=J\sum_n\left(\begin{array}{c}
C_n \\
a_{2n+1}C_n \\
a_{2n+1}D_n \\
D_n
\end{array}\right)
e^{-i[\eps+(2n+1)eV]t}\\[-0.6cm]\nn
\eeqa
\end{subequations}
to the left ($x=0$) and the right ($x=L$) of the junction, respectively.
Here, $a_n(\eps)=a(\eps+neV)$, where
\beq
a(\eps)=\frac1\Delta\left\{\begin{array}{ll}
\eps-i\sqrt{\Delta^2-\eps^2}, & |\eps|<\Delta,\\
\eps-\sign(\eps)\sqrt{\eps^2-\Delta^2}, & |\eps|\geq\Delta,
\end{array}\right.
\eeq
is the Andreev reflection amplitude, and the prefactor $J(\eps)=\sqrt{1-|a(\eps)|^2}$ is set by the normalization of the wavefunction.

The exchange field $M$ leads to backscattering of right movers into left movers (and vice versa).
This process is described by unitary scattering matrices 
\beq
S_e=\left(\begin{array}{cc}
r & t  \\
t & -r^*t/t^*
\end{array}\right)
\;\mathrm{and}\enspace
S_h=\left(\begin{array}{cc}
-r^* & t^*  \\
t^* & rt^*/t
\end{array}\right)
\eeq 
for electrons and holes, respectively, that  are
characterized by a
transmission probability 
$D=|t|^2$. 
In the short junction limit $L\ll v/\Delta$ -- where $L$ is the distance between the leads -- and assuming $M\gg \Delta$, one obtains $D=[1+\sinh^2(LM/v)]^{-1}$  \cite{fu2009}. 

The scattering matrices relate the coefficients $A_n,B_n,C_n,D_n$ through the set of equations
\begin{subequations}
\beqa
\left(\begin{array}{c}
B_n \\
C_n
\end{array}\right)
&=&
S_e
\left(\begin{array}{c}
\delta_{n,0}+a_{2n}A_n\\
a_{2n+1}D_n \\
\end{array}\right),
\\
\left(\begin{array}{c}
A_n \\
D_{n-1}
\end{array}\right)
&=&
S_h
\left(\begin{array}{c}
a_{2n}B_n\\
a_{2n-1}C_{n-1} \\
\end{array}\right),
\eeqa
\end{subequations}
that can be solved numerically \cite{averin1995}. 

Assuming that the incoming quasiparticles states are occupied according to the equilibrium distribution at temperature $T$ in the leads, the average current $I(t)=\langle \hat I(t) \rangle$ in the stationary regime \cite{short-times} takes the form 
\begin{eqnarray}
I(t)=\sum_n I_n e^{i2neVt},
\label{eq-I_n}
\end{eqnarray}
where
\begin{widetext}
\vspace*{-0.3cm}
\beqa
I_n&=&
\frac eh
\left\{
DeV\delta_{n0} 
-\!\int\! d\eps\; \tanh\frac{\eps}{2T}
J^2
\left[a^*_{2n}A^*_n+a_{-2n}A_{-n}
+
\sum_m(1+a^*_{2(m+n)}a_{2m})
\left(A^*_{m+n}A_m-B^*_{m+n}B_m\right)
\right]\right\}.
\label{eq:current}
\eeqa
\end{widetext}
In Fig.~\ref{fig1}, we plot the numerical 
results at $T=0$ for the dc current $I_0(V)$ as well as the real and imaginary part of the first harmonic $I_1(V)$ (at the conventional Josephson frequency).

Let us first discuss the dissipative current $I_0(V)$. For a transparent junction ($D=1$), the current flowing through the junction is exactly half of the current of a perfectly transmitting conventional Josephson junction. This is not surprising as the BdG Hamiltonians for both systems are identical in the absence of backscattering -- however, in conventional junctions, there are two copies of that Hamiltonian. As $V\rightarrow 0$, the dissipative current reaches the same value $(2/\pi)I_c$, where $I_c=(e/2\hbar)\Delta$ is the critical current of the junction. 

At finite backscattering ($D<1$), we find that $I_0$ vanishes in the limit $V\rightarrow 0$. This current suppression can be understood due to the energy gap $E_{\rm g}=(1-\sqrt{D})\Delta$ between the bound states and the continuum. In addition, singularities appear at voltages $eV=\Delta/q$, where $q$ is an integer~\cite{cuevas-fogelstrom}. This is to be contrasted with conventional Josephson junctions, where singularities appear at voltages $eV=2\Delta/q$.  These singularities are associated with the energy gap $2\Delta$ between occupied and empty states that quasiparticles have to overcome. Namely, new channels for charge transfer open at the specific voltages $eV=2\Delta/q$ when quasiparticles can overcome the energy gap by performing $q-1$ Andreev reflections. In S/TI/S junctions,  the presence of the MBS 
reduces the energy gap between occupied and empty states to $\Delta$. 

This manifests itself most clearly in the tunneling regime ($D\ll1$), where one sees a current onset at $eV=\Delta$, cf. Fig.~\ref{fig1}. The analytic expression for the dc current in the tunneling regime reads
$
I_0^\mathrm{tun}(V)=
(eD/h)
\int d\eps\,
\nu(\eps)\nu(\eps-eV)[f(\eps-eV)-f(\eps)]
$.
Here, $f(\eps)$ is the Fermi distribution, and the normalized density of states on either side of the junction,
$
\nu(\eps)
=
\pi \Delta \delta(\eps)+\theta(|\eps|-\Delta){\sqrt{\eps^2-\Delta^2}}/{|\eps|}
$,
shows the contributions of MBS and continuum.
At $T=0$, the current $I_0^\mathrm{tun}$ is the sum of two terms. The first term corresponds to the transfer from the continuum to the MBS for voltages $eV\geq\Delta$, whereas the second term corresponds to the (conventional) transfer from continuum to continuum for voltages $eV\geq2\Delta$. The suppression of the square-root singularity at the gap edge in $\nu(\eps)$ explains that the singular behavior of $I_0(V)$ at $eV=2\Delta$ is smooth. 

For completeness, we provide  the analytical expression for the excess current,
$I_{\rm exc}=I_0-(e^2/h)DV$
 at $eV\gg\Delta$,
$
I_{\rm exc}
=
({2 e}/h)({D^2\Delta}/{R})(1-(D/{\sqrt R})\arctan\sqrt{R})$,
where $R=1-D$, which varies between $0$ in the tunneling limit and $(8/3\pi)I_c$ at perfect transmission.

We now turn to the ac components of the current. Multiple Andreev reflections yield ac components at multiples of the Josephson frequency $\omega_J$. In particular, concentrating on the first harmonic $I_1$, we notice that, just as the dc current, both its real and imaginary part show MAR features at $eV=\Delta/q$. More strikingly, $I_1$ vanishes in the limit $V\to0$ (except for perfectly transparent junctions), and in the tunneling limit, $I_1$ is strongly suppressed even at finite voltages. 

At first sight, these results seem somewhat paradoxical: Does the vanishing of $I_1$ indicate the absence of an ac Josephson effect? In fact, the fractional Josephson effect with frequency $\omega_J/2$ is absent in the formalism from the outset, cf. Eq.~\eqref{eq-I_n}, 
whereas the regular Josephson effect with frequency $\omega_J$ 
vanishes at small voltage. The latter is in stark contrast with the case of conventional Josephson junctions, where the stationary value of the dc current is recovered as $V\to 0$ \cite{averin1995}. 

To resolve this puzzle, we study the current-current correlations. In particular, we consider the (symmetrized) frequency-dependent current noise,
\beq
S(\omega)
=\int d\tau
e^{i\omega \tau}
\overline{
\langle \delta \hat I(t) \delta \hat I(t\!+\!\tau)+\delta \hat I(t\!+\!\tau)\delta \hat I(t)\rangle
}, 
\eeq
where $\delta \hat I(t)=\hat I(t)-I(t)$, and the bar denotes a time averaging. In terms of the coefficients $A_n, B_n, C_n, D_n$, the noise can be expressed in the following form,
\begin{widetext}
\beqa
S(\omega) &=&
\frac{e^2}{2h}
\sum_{ \pm \omega } 
\sum_p
\int d\eps\, d\eps'\;
J(\epsilon)J(\epsilon')
\Bigg\{
\left(f(\eps)[1-f(\eps')]+f(\eps')[1-f(\eps)]\right)\times
\\
&&
\quad\times
\Bigg[
\delta(\eps-\eps'  \pm \omega  + 2peV)
\Big(
\left|
\sum_n
\left[
A_{n + p}^*A{'_n} + a_{2(n + p)}^*a{'_{2n}}\bar A_{n + p}^*\bar A{'_n} - (1 + a_{2(n + p)}^*a{'_{2n}})B_{n + p}^*B{'_n}
\right]
\right|^2
\nn\\
&&
\qquad\qquad\qquad\qquad\qquad\qquad
+
\left|
\sum_n
 (1 + a_{2(n + p) + 1}^*a{'_{2n + 1}})
 \left( {C_{n + p}^*C{'_n} - D_{n + p}^*D{'_n}} \right)
\right|^2
\Big)
\nn\\
&&
\quad
+
2\delta(\eps-\eps'  \pm \omega  + (2p-1)eV)
\left|
\sum_n
\left[A_{n + p}^*C{'_n} + a_{2(n + p)}^*a{'_{2n + 1}}\bar A_{n + p}^*C{'_n} -(1 + a_{2(n + p)}^*a{'_{2n + 1}})B_{n + p}^*D{'_n} 
\right]
\right|^2
\Bigg]
\nn\\
&&
+
\left(f(\eps)f(\eps')+[1-f(\eps)][1-f(\eps')]\right)\times
\nn\\
&&
\quad\times
\Bigg[
\delta(\eps+\eps' \pm \omega  + 2peV)
\left|
\sum_n
\left[{{A_{p - n}}B{'_n} -{B_{p - n}}A{'_n} - {a_{2(p - n)}}a{'_{2n}}\left( {{{\bar A}_{p - n}}B{'_n} -{B_{p - n}}\bar A{'_n}} \right)} 
\right]
\right|^2
\nn\\
&&
\quad+\delta(\eps+\eps' \pm \omega  + 2(p+1)eV)
\left|
\sum_n
(1 - {a_{2(p - n) + 1}}a{'_{2n + 1}})\left( {{D_{p - n}}C{'_n}-{C_{p - n}}D{'_n}} \right)\right|^2
\nn\\
&&
\quad+2\delta(\eps+\eps' \pm \omega  + (2p+1)eV)
\left|
\sum_n
\left[
{{A_{p - n}}D{'_n} - {a_{2(p - n)}}a{'_{2n + 1}}{{\bar A}_{p - n}}D{'_n} - \left( {1 - {a_{2(p - n)}}a{'_{2n + 1}}} \right){B_{p - n}}C{'_n}}
\right]
\right|^2
\Bigg]\Bigg\}.\nn
\eeqa 
\end{widetext}
Here, 
$\bar A_n\!=\!A_n\!+\!\delta_{n0}/a$, and all unprimed (primed) quantities are functions of $\eps$ ($\eps')$.

\begin{figure}
\includegraphics[width=0.85\linewidth]{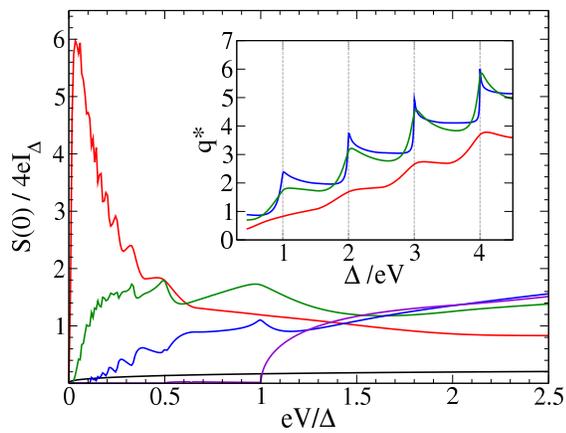}
\caption{
Zero-frequency current noise for the same transparencies as in Fig. \ref{fig1}. The inset shows the effective charge $q^*$  as a function of applied bias for $D=0.4$, $0.7$, and $0.9$.
}
\label{fig2} 
\vspace*{-0.3cm}
\end{figure}

We first analyze the zero-frequency noise which is shown in Fig.~\ref{fig2}. Taking the ratio of the noise and the dc current, one can define an effective charge $q^*=S/(2I_0)$. As shown in the inset of Fig.~\ref{fig2}, the effective charge is $q^*\sim \Delta/V$, which has to be compared to $q^*\sim 2\Delta/V$ in a conventional Josephson junction \cite{cuevas1999,naveh1999}. This result is directly related to the modified positions of the MAR features in the dc current discussed above.

\begin{figure}
\includegraphics[width=0.85\linewidth]{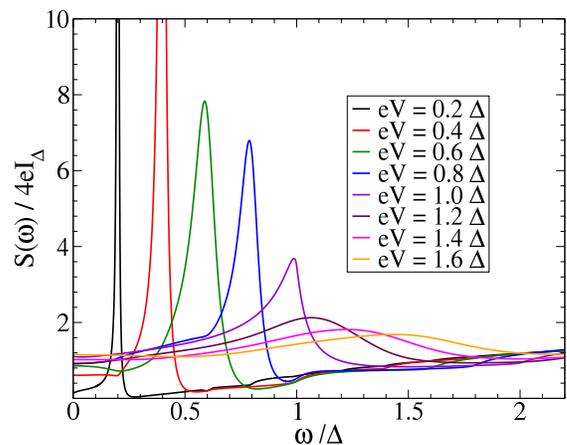}
\caption{
Finite frequency noise of a junction with transparency $D=0.4$  for various values of the applied bias.
}
\label{fig3} 
\vspace*{-0.3cm}
\end{figure}

More interesting is the finite-frequency noise which is shown in Fig.~\ref{fig3} for various values of the applied bias. The most striking feature is a peak at $\omega=eV$. The peak is revealed in the analytical formula for the noise
$S^\mathrm{tun}(\omega)=e\sum_\pm I_0^\mathrm{tun}(V\pm\omega/e)\coth[(eV\pm \omega)/2T]$
in the tunneling regime ($D\ll 1$); it is absent at $D=1$.
In a conventional Josephson junction, the peak is absent at any transmission. Here it is a consequence of the $4\pi$-periodicity of the Andreev bound states. This can be understood in the following way. The two Andreev bound states carry a current $I_\pm=\pm(e/2\hbar)\sqrt{D}\Delta\sin(\varphi/2)$. If the system would remain in one of the two states indefinitely, one would indeed observe a fractional Josephson effect. However, due to the coupling between the bound state and the continuum in the presence of an applied bias, the switching probability is non-zero. As a consequence, the fractional Josephson effect disappears in the average current, but its signature remains visible in the noise spectrum.  
In particular, in the limit $V\to0$, where the switching probability is infinitesimal, the system is equally likely to be in either state and the current averages  to zero. At finite bias, the switching probability is finite, and 
a conventional $2\pi$-periodic effect remains  \cite{unpublished}. 

To summarize, we studied the out-of-equilibrium properties of S/TI/S junctions. The dc current shows clear signatures of the Majorana bound states forming in the junction. The expected fractional Josephson effect with frequency $\omega_J/2$ is absent in the average current as the applied bias introduces relaxation processes. However, its signatures can be seen in the finite-frequency noise which displays peaks at $\omega=eV$. We believe that our results are directly applicable to the S/TI/S junction studied in Ref.~\cite{knez2011b}, where the absence of a supercurrent signature in the differential resistance was attributed to
the lack of coherence in the junction. Frequency-dependent noise experiments \cite{freq-noise}  could reveal whether coherent effects do exist in this sample.

We would like to acknowledge helpful discussions with Leonid Glazman and Gil Refael. Part of this research was supported through a Marie Curie IRG and the Fondation Nanosciences of Grenoble. Furthermore, JM thanks the Aspen Center for Physics for hospitality.


\begin{thebibliography}{99}
\bibitem{otheredge}
Note that we consider only one edge. The edge states on the other edge of the TI may either be suppressed, e.g., by applying a large magnetic field or, if the two edges are sufficiently far apart, their contribution would add incoherently to the result.
\end{thebibliography}
\end{document}